\documentstyle[a4,11pt,epsf]{article}
\epsfverbosetrue
 0

\newcommand{\beq}{\begin{equation}}
\newcommand{\eeq}{\end{equation}}

\begin{document}

\begin{titlepage}
 


\vspace{5mm}
 
\begin{center}
{\huge 
       The Weakly Coupled Gross-Neveu 
       \\[3mm]
       Model with Wilson Fermions
}\\[15mm]
{\bf R. Kenna and J.C. Sexton,
\\
School of Mathematics, Trinity College Dublin, Ireland} 
\\[3mm]~\\ 
March 2001
\end{center}
\begin{abstract}
The nature of the phase transition in the lattice
Gross-Neveu model with Wilson fermions is investigated
using a new analytical technique.
This involves a new type of weak coupling expansion
which focuses on the partition function zeroes of the model.
Its application to the single flavour
Gross-Neveu model yields a phase diagram
whose structure is 
consistent with that predicted from a saddle point approach.
The existence of an Aoki phase is confirmed
and its  width  in the weakly coupled
region is determined.
Parity, rather than chiral symmetry breaking 
naturally emerges as the 
driving mechanism for the phase transition.
\end{abstract}
\end{titlepage}

\newpage

\section{Introduction}
\setcounter{equation}{0}
In continuum QCD, the conventional explanation for the smallness of
the mass of pseudoscalar $\pi$ mesons is the following: QCD with
$N_f$ massless quark flavours has a  global chiral 
$U(N_f) \times U(N_f)$ symmetry, which, spontaneously broken, reduces
to $U(N_f)$ and yields $N_f^2$ Goldstone bosons. Explicit 
breaking of the original chiral symmetry by a small quark mass renders
these $N_f^2$ Goldstone bosons massive with correspondingly small 
mass. To accord with nature, one of these Goldstone bosons (the 
$\eta$-particle in the case $N_f=2$) has to acquire an an additional
mass. To explain this was known as the $U(1)$ problem in the continuum.
Its resolution there comes from the axial anomaly, whereby  the 
axial symmetry corresponding to the $U(1)$ subgroup of $U(N_f)$ is 
explicitly broken by a quantum  effect, reducing the number of 
Goldstone bosons to $N_f^2 -1$.

Naive lattice regularization of such a fermionic theory is
hindered by the doubling problem, namely that a return to the continuum
manifests too many fermionic degrees of freedom. This doubling problem
is resolved by the usage of Wilson fermions. However the extra Wilson 
term that removes the fermion doublers breaks chiral symmetry explicitly. 
This effect can be traced back to the existence of the 
axial anomaly in the continuum. For this reason the staggered 
format has  often been the favoured one for the  study of models
with chiral symmetry breaking \cite{HaKo93}.

The Wilson action for free fermions 
in terms of dimensionless fermionic fields $\psi (n)$ defined at the
sites $n$ of a $d$-dimensional lattice is
\begin{equation}
S_F^{(0)}[ 
          \bar{\psi},\psi 
         ]
 =
\frac{1}{2\kappa}
\sum_n{ 
       \bar{\psi}(n) \psi (n)
      }
-
\frac{1}{2}
\sum_{n,\mu}{ 
             \left[
                    \bar{\psi}(n)     (r - \gamma_\mu) \psi (n + \mu)
                +   \bar{\psi}(n+\mu) (r + \gamma_\mu) \psi (n)
             \right]
            }
\quad ,
\label{cfRo14.2}
\end{equation}
where 
\beq
 1/2 \kappa =  \hat{M}_0 + dr
\quad .
\label{kappa}
\eeq
Here, $\kappa$ is the hopping parameter, $r$ is the Wilson parameter, $a$
is the lattice spacing and $\hat{M}_0$ is a dimensionless 
fermion bare mass parameter. We use $d=2$ and $r=1$ throughout.
This free fermion model is the weak coupling limit of an interactive 
theory in which a bare parameter $g$ measures the coupling 
of the free theory to some interaction.
Even if $\hat{M}_0=0$, now, the Wilson term contributes to the hopping 
parameter and there is no obvious chiral symmetry. The question arises
- what is the status of the chiral phase transition and the $U(1)$ 
problem on the lattice?

Despite the lack of an obvious chiral symmetry,
there exists a host of numerical and analytical evidence for the 
existence of massless pions in the lattice formulation of QCD. These are 
believed to exist on a critical line $\kappa_c(g)$.
In the literature, there are  two explanations
for the existence of the critical line and the masslessness of the lattice
pions. 

The first of these was sometimes referred to as the conventional
explanation \cite{Kawamoto}. 
Although there is no obvious chiral symmetry at non-zero $r$,
the conventional explanation suggests that tuning $\kappa$ effects its 
recovery in some unknown way. Now, with chiral symmetry recovered at 
$\kappa_c(g)$, the same arguments as in the continuum may be applied.

The second explanation was first forwarded in 1984 by Aoki 
\cite{Aoki}.
Here it is accepted that since there is no chiral symmetry in the lattice
formulation of QCD, its spontaneous breaking cannot be responsible for 
the masslessness of pions. Instead there is an Ising-like second order
parity breaking phase transition. In the single flavour
case
the order parameter for parity symmetry is $\bar{\psi} i \gamma_5 \psi$,
the operator corresponding to the single $\pi$ meson. The parity symmetric 
phase is where 
$\langle \pi \rangle = \langle \bar{\psi} i \gamma_5 \psi
\rangle = 0 $. there is also a phase with long range order where 
$\langle \pi \rangle \ne 0$. At the transition between these phases, 
a correlation length $\xi$ diverges. This correlation length is identified 
as the inverse of the pion mass, which, hence, becomes zero on the 
phase boundary. Thus the pion is not a Goldstone boson in the Wilson 
lattice 
formulation. Aoki also recovered the current algebra relation between 
pion and quark  mass ($m_\pi^2 \sim m_q \sim (\kappa - \kappa_c)$) by 
considering the effective meson theory as a scalar field theory in four 
dimensions with mean field like critical behaviour.
In the multiflavour case the parity symmetry breaking is accompanied by a 
flavour symmetry breaking and, with it, Goldstone bosons in the form
of the charged pions. The $\eta$ remains massive according to 
Aoki's analysis, and the $U(1)$ problem in the lattice successfully 
resolved \cite{Aoki,Ao86}.

Two main features distinguish Aoki's QCD phase diagram from the 
conventional one. Firstly, the existence of
the phase transition in Aoki's scenario is due to parity 
symmetry breaking as opposed to chiral symmetry breaking in the 
conventional picture. The order parameter is 
$\langle \bar{\psi} i \gamma_5 \psi \rangle$ 
rather than 
$\langle \bar{\psi} \psi \rangle$ \cite{Ao86}.
Secondly, instead of a single critical line 
extending from the strongly coupled limit $g = \infty$ to $\kappa = 
1/2d$ in the weakly coupled limit $g=0$, Aoki's picture involves
the existence of two such lines 
and (in QCD) five critical points linked by four cusps in the weakly coupled
zone.

Aoki's QCD phase diagram is based on infinite-volume analyses in the limits
of strong and weak coupling and on an analogy to the Gross-Neveu model,
which, except for confinement,  has features similar to QCD. One of
these features is asymptotic freedom, so that in the Gross-Neveu model,
as in QCD, the continuum limit is taken in the weakly coupled regime.
Aoki's scenario in the Gross-Neveu model again involves two critical
lines spanning the full coupling range, with three critical points
at zero coupling, linked by two cusps. 
This picture is based on saddle point methods \cite{Aoki}.
There exists substantial evidence in support of this scenario in the 
strongly
coupled regime 
\cite{Aoki,Ao86,support1,support2,AoBo94,BiVr95b,numericsandreview,Bitarsupport,width,Sharpe,us}. 
In the weakly coupled regime, however, the evidence has  been 
clear cut \cite{BiVr95,against} 
and this is the region where our
attention is focused.
Recently, also, Creutz \cite{Creutz} has posed a question
as to the size of the Aoki phase. This question is
 whether
the Aoki phase
is ``squeezed out'' between the arms of the cusps
at non-zero coupling or whether
 it only vanishes in the weak coupling limit \cite{width,Sharpe}.

This sets the twofold motivation for this paper. 
Firstly, a  new type of
weak coupling expansion is developed \cite{us}.
From it,  the partition function zeroes
of Wilson fermionic  models can be extracted in a natural way. 
This weak coupling technique is then applied 
to the Gross-Neveu model, where the existence of an Aoki
phase was first suggested \cite{Aoki}.
We confine our attention to the single flavour Gross-Neveu model
and variants thereof. 
We also address the question of the ``squeezing out'' of
the Aoki phase at weak coupling.
This multiplicative approach to the single flavour Gross-Neveu model,
shows that the width of the central  Aoki cusp
is   ${\cal{O}}({g}^2)$
while the Aoki phase has not yet emerged at this order
from the left and right extremes.
Furthermore, that parity symmetry
breaking is the phase transition mechanism emerges in a very transparent 
way.

\section{The Gross-Neveu Model}
\setcounter{equation}{0}

The original motivation for the introduction of the Gross-Neveu model
in the continuum 
\cite{GN} was to study a renormalisable quantum field theory involving 
dynamical spontaneous symmetry breaking. Such models evolved from
four dimensional four-fermi models studied by Nambu and Jona-Lasinio
\cite{NJL} and are essentially their two dimensional equivalents. The 
Gross-Neveu model is, however, renormalisable and asymptotically free.
It is a model of fermions only, which interact
through a short range quartic interaction.
We start with a generalized Gross-Neveu model, whose action, in
euclidean continuum space, is given by 
\begin{equation}
 S_{ \rm{GN} }^{ \rm{(cnm)} }
=
\int{
  d^2x 
     \left\{
             {
                     \bar{\psi} (x) 
                     \left( 
                           \partial \!\!\!/ 
                           \:+ M 
                     \right) 
                     \psi (x)
                   }
                   -
                   \frac{g_\phi^2}{2}
                         \left(
                                    {
                                      \bar{\psi}(x) \psi(x) 
                                    }
                         \right)^2
                   -
                   \frac{g_\pi^2}{2}
                         \left(
                                     {
                                      \bar{\psi}(x)  i \gamma_S \psi(x) 
                                     }
                         \right)^2
      \right\}
}
\quad ,
\label{cfIz2.1cfHa1.1}
\end{equation}
where $\gamma_S = i^{-1}\gamma_1 \gamma_2$
and $\psi(x)$ is a $2$  component fermion field.
Note that we have allowed for two different four-fermion couplings.
This allows for some flexibility
to tune in or out the continuous chiral symmetry
present in the continuum action \cite{splitgsAoki,splitgs}.

We use the following representation for the Dirac $\gamma$-matrices
in two dimensions,
\begin{equation}
 \gamma_1^{(d=2)} = 
 \left( \begin{array}{cc}
 0 & 1 \\
 1 & 0
 \end{array} \right)
\quad \quad \quad \quad 
 \gamma_2^{(d=2)} = 
 \left( \begin{array}{cc}
 0 & -i \\
 i & 0
 \end{array} \right)
\quad ,
\label{5.1}
\end{equation}
so that the chirality operator is
\begin{equation}
\gamma_S 
 =  
 i^{-1}
 \gamma_1  \gamma_2
 =  
 \left( \begin{array}{cc}
                         1 & 0 \\
                         0 & -1
        \end{array}
 \right)
\quad .
\label{ZJp109}
\end{equation}

Each term in the action (\ref{cfIz2.1cfHa1.1}) is invariant under the 
continuous global $U(1)$ symmetry
\begin{equation}
 \psi(x) \rightarrow \exp{(i \alpha)} \psi (x)
\quad ,
\quad \quad \quad \quad \quad \quad \quad \quad 
 \bar{\psi}(x)  \rightarrow \exp{(-i \alpha)} \bar{\psi}(x)
\quad .
\label{cnssymm1}
\end{equation}

If, further, the fermion mass $M$ vanishes, the action 
(\ref{cfIz2.1cfHa1.1}) is also invariant under a discrete global chiral
$Z_2$ transformation
\begin{equation}
 \psi(x) \rightarrow \gamma_S \psi(x)
\quad ,
\quad \quad \quad \quad \quad \quad \quad \quad 
 \bar{\psi  }(x)   \rightarrow \bar{\psi}(x) \gamma_S
\quad .
\label{discretechisymm}
\end{equation}
This is the symmetry of the original (standard) 
version of the model, in which the last term of 
(\ref{cfIz2.1cfHa1.1}) is absent (i.e., $g_\pi = 0$).
Finally, if the four fermi couplings are tuned such that
$g_\phi = g_\pi$, the discrete chiral symmetry is promoted to a continuous 
one, namely,
\begin{equation}
 \psi(x) \rightarrow \exp{(i \theta \gamma_S)} \psi(x)
\quad ,
\quad \quad \quad \quad \quad \quad \quad \quad 
 \bar{\psi}(x)   \rightarrow \bar{\psi}(x) \exp{(i \theta \gamma_S)}
\quad .
\label{cnssymm2}
\end{equation}
This cannot  be broken since there are no Goldstone bosons in two dimensions
due to the Mermin-Wagner theorem \cite{MW}. Nonetheless, a 
topological long range order of the Kosterlitz-Thouless type
could exist in the model \cite{KT-KeIr96}.
The Mermin-Wagner theorem refers only to continuous symmetries and
does not preclude the spontaneous breaking of a discrete symmetry in
two dimensions. In the continuum
Gross-Neveu model, the spontaneous breaking
of the discrete $\gamma_S$ symmetry leads to dynamical fermion
mass generation.
The mass term explicitly breaks chiral symmetry and is
analogous to an
external field in the Ising model, say.

Bosonizing the action gives for the partition function,
\begin{equation}
 Z_{GN}^{({\rm{cnm}})} =
\int{
 {\cal{D}} \phi
 {\cal{D}} \pi 
 {\cal{D}} \bar{\psi}
 {\cal{D}} \psi
 e^{-S}
}
\quad ,
\end{equation}
where
\begin{equation}
 S
=
\int{
     d^dx \left\{
                 \bar{\psi}(x) \left( \partial \!\!\!/ 
                         \:+ M \right) \psi(x)
                +
                \frac{1}{2g_\phi^2}
                \phi^2(x)
                +
                \frac{1}{2g_\pi^2}
                \pi^2(x)
               +
                     \phi(x)
               \bar{\psi}(x) \psi(x)
               +
               \pi(x)
               \bar{\psi}(x) i \gamma_S \psi(x)
       \right\}
}
\quad ,
\label{S}
\end{equation}
where  $\phi(x)$ and $\pi(x)$ are auxiliary boson fields. 
The chiral transformations now represent rotations
between these auxiliary fields.

\section{Lattice Regularization with Wilson Fermions}
\setcounter{equation}{0}

Lattice regularization of the bosonized Gross-Neveu model, with Wilson
fermions, leads to the action
\begin{equation}
 S_F^{  ({\rm{W}})  } [\phi, \pi, {\bar{\psi}}, \psi ]
  =  S_F^{(0)}[{\bar{\psi}}, \psi ] + 
S_{  ({\rm{int}})  } [\phi, \pi, {\bar{\psi}}, \psi ]
+ S_{  ({\rm{bosons}})  }[\phi, \pi]
\quad ,
\end{equation}
where \cite{support1}
\begin{equation}
 S_F^{(0)}
 =
 \frac{1}{2\kappa}
 \sum_n{
        \bar{\psi}(n)\psi(n)
       }
 -
 \frac{1}{2}\sum_{n,\mu}{
                           \left\{
                                  \bar{\psi}(n)
                                  ( 1 - \gamma_\mu )
                                 \psi(n+\hat{\mu})
                                 + 
                                 \bar{\psi}(n+\hat{\mu})
                                 ( 1 + \gamma_\mu )
                                 \psi(n)
                           \right\}
                         }
\quad ,
\label{cfRo14.2d}
\end{equation}
\begin{equation}
 S_{  ({\rm{int}})  } 
=
 \sum_n{\phi(n) \bar{\psi}(n) \psi(n)
}       
+
 \sum_n{\pi(n) \bar{\psi}(n) i \gamma_S \psi(n)
}       
\quad,
\end{equation}
and
\begin{equation}
 S_{  ({\rm{bosons}})  } 
=
 \frac{1}{2{{g}}_\phi^2 }
 \sum_n{ \phi^2(n) }       
+
 \frac{1}{2{{g}}_\pi^2 }
 \sum_n{ \pi^2(n) }       
\quad.
\label{pure}
\end{equation}
The lattice sites are labeled 
$ n_\mu = -N/2, \dots, N/2 - 1$, and $N$ is the 
 number of sites in each of the two directions, which 
we assume to be even.
Appropriate tuning of the two couplings $g_\phi^2$ and $g_\pi^2$
may allow recovery of chiral symmetry in the continuum limit 
(see \cite{splitgs} for discussions).

 Lattice Fourier 
transforms are defined as
\begin{equation}
 f(n) = 
 \left( \frac{1}{Na} \right)^2 \sum_{p}{
 \tilde{f}(p) e^{i p.na}} \quad,
\quad \quad \quad 
 \tilde{f}(p) = a^2 \sum_{n}{
 f(n) e^{-i p.na}} \quad,
\label{1.1}
\end{equation}
where
\begin{equation}
 p_\mu = \frac{2\pi}{Na} {\hat{p}}_\mu
\quad ,
\label{1.3}
\end{equation}
and where ${\hat{p}}_\mu$ are  integers or half integers depending
on the field type and the boundary conditions. We henceforth drop
the tilde on Fourier transformed field variables.

The fermionic part of the action can be expressed 
in terms of momentum space variables as
\begin{equation}
 S_F^{(\rm{0})} [\bar{\psi},\psi] 
+ S_{\rm{(int)}} [\phi,\pi,\bar{\psi},\psi] 
 = 
 \frac{1}{a^{4}}
 \frac{1}{N^2}
   \sum_{q,p}{
                           \bar{\psi}(q)  
                           M^{(W)} (q,p) \psi (p)
                         }
\quad.
\label{3.1}
\end{equation}
Here  $M^{(W)} (q,p)$ are  $2\times 2$ matrices and
\begin{equation}
 M^{(W)}(q,p) = M^{(0)}(q,p) + M^{({\rm{int}})}(q,p)
\quad,
\label{3.2}
\end{equation}
with
\begin{eqnarray}
 M^{(0)}(q,p) & = & \delta_{p,q} M^{(0)}(p)
\quad , \label{3.3} \\
 ~   & = & 
 \delta_{p,q} \left[
 \frac{1}{2\kappa}
 - \sum_\mu{
              \cos{p_\mu a}
            } 
 + i \sum_\mu{
             \gamma_\mu \sin{p_\mu a}
            }
 \right]
\quad ,
\label{3.4}
\end{eqnarray}
and
\begin{equation}
M^{({\rm{int}})}(q,p)
=
\frac{1}{N^2}
\sum_n{
e^{i (p-q) n a}
 \left[
  \phi(n) + \pi (n) i \gamma_S
 \right]
}
\quad .
\label{3.12}
\end{equation}
Integration over the Grassmann variables gives the full
partition function 
\begin{equation}
Z = 
\int{
 {\cal{D}} \phi
 {\cal{D}} \pi 
 {\cal{D}} \bar{\psi}
 {\cal{D}} \psi
 \exp{\left(-S_F^{\rm{(W)}}\right)}
}
\propto \left\langle 
 \det{  M^{  ({\rm{W}})  }  } \right\rangle
\propto \left\langle 
  \prod_{\alpha,p} \lambda_\alpha(p)  \right\rangle
\quad ,
\label{zzzz}
\end{equation}
with $\lambda_\alpha(p)$ the eigenvalues of the fermion matrix
and the expectation
values being taken over the bosonic fields.
Note that there is no hopping parameter dependence in $M^{({\rm{int}})}$.

In the free fermion case the partition function is simply 
proportional to 
\begin{equation}
 \det{M^{(0)}} = \prod_{\alpha,p} \lambda_\alpha^{(0)}(p)
\quad,
\label{freedet1}
\end{equation}
where $\lambda_\alpha^{(0)}(p)$ are the eigenvalue solutions of
\begin{equation}
 M^{(0)}(p) | \lambda^{(0)} \rangle
 =
 \lambda^{(0)} | \lambda^{(0)} \rangle
\quad .
\label{5.5}
\end{equation}
Using the representation (\ref{5.1}) for the Dirac $\gamma$-matrices,
the solution to this problem is easily found to be
\begin{eqnarray}
| \lambda^{(0)}_\alpha (p) \rangle 
 & = &
 \frac{1}{\sqrt{2}}
 \left( \begin{array}{c}
 1 \\
 (-1)^\alpha \frac{\sin{p_1a}+i\sin{p_2 a}}{\sqrt{
  \sum_{\mu=1}^2{\sin^2{p_\mu a}}}}
 \end{array} \right)
\quad,
\label{5.8}
\\
 \lambda^{(0)}_\alpha (p) 
 & = &
 \frac{1}{2\kappa}
 - \sum_{\mu=1}^2 \cos{p_\mu a} 
 + i (-1)^\alpha  \sqrt{ \sum_{\mu=1}^2{\sin^2{p_\mu a}}}
\quad,
\label{5.7}
\end{eqnarray}
 where $\alpha=1,2$.
These eigenfunctions form a complete orthonormal set.
 As is usual for Grassmann variables,
we impose antiperiodic boundary conditions in the
temporal ($1$-) direction
and periodic boundary conditions
in the spatial ($2$-) direction. 
With these mixed boundary conditions the momenta ${\hat{p}}_\mu$
for the Fourier transformed fermion fields
take the integer or half-integer values,
$ {\hat{p}}_1 = -N/2+1/2,  \dots, N/2 - 1/2$ and
$ {\hat{p}}_2  =  -N/2,  \dots, N/2 - 1$.
Then, the eigenvalues (\ref{5.7}) in the free fermion
case are either two-fold 
or four-fold 
degenerate, the former being the case if
$\hat{p}_2 = 0$ or $-N/2$.

In the free fermion case the Lee-Yang Zeroes \cite{LY} are given by
$\lambda_\alpha^{(0)}(p) = 0$. From (\ref{5.7}), this is the case at 
\begin{equation}
 \frac{1}{2\kappa} 
 =
 \eta_\alpha (p) =
 \sum_{\mu=1}^2 \cos{p_\mu a} 
 - i (-1)^\alpha  \sqrt{\sum_{\mu=1}^2{\sin{p_\mu a}}}
\quad.
\label{20.1}
\end{equation}
The lowest zeroes (with the smallest imaginary parts) correspond
to 
\begin{equation}
\hat{p} = (\pm (N/2-1/2),-N/2) \quad,\quad
(\pm 1/2, -N/2) \quad,\quad
(\pm (N/2-1/2),0)\quad  {\rm{and}} \quad
(\pm 1/2,0)
\quad ,
\label{list}
\end{equation}
impacting onto the real axis at $1/2\kappa = -2, 0, 0$ and $2$ 
respectively. These are precisely the three nadirs of the Aoki cusps
in the Gross-Neveu model (see Fig.~1).

Note that the zeroes in the upper half plane
are given by $\alpha = 1$, while their complex conjugates correspond
to $\alpha = 2$.
Note, further,
 that the zeroes in (\ref{20.1}) are two or four fold degenerate
in the momenta. I.e., these zeroes are invariant under $p_\mu 
\rightarrow - p_\mu$. 
This transformation is just a rotation through an angle $\pi$
in the space-time plane.
The lowest zeroes (\ref{list}), which are responsible for the 
critical behaviour of the free model, are 
actually two-fold degenerate.
There is also a symmetry under $p_1
\leftrightarrow  p_2$ which is manifest in the infinite volume
limit. This is equivalent to a trivial rotation by $\pi/2$ in the
$p_1$-$p_2$ plane, followed by reflection through the $p_2$ axis.
Since this reflection is through the spatial axis, this transformation
is, in fact, parity. I.e., apart from rotations in space and time, 
the critical points are left unchanged under
the parity transformation.

\section{A New Weak Coupling Expansion}
\setcounter{equation}{0}

The usual weak coupling expansion of the full 
determinant for a general fermionic field
theory is the Taylor expansion of
\begin{eqnarray}
 \det{ M^{ ( { \rm{W} } ) } }
 &=&
 \det{ M^{(0)} }
 \times 
 \det \left({ M^{(0)} }^{-1}
 M^{ ( {\rm{W}} ) }\right)
\nonumber \\
~ & = &
 \det{ M^{(0)} }
  \exp{ 
       {\rm{tr}}
       \ln{
            \left(
                  1+{M^{(0)}}^{-1}
                    M^{ ( { \rm{int} } ) }
            \right)
          }
      }
\quad .
\end{eqnarray}
This expansion is additive in nature and, from it,
the ratio of full to free fermion determinants
may be written,
\begin{equation}
 \frac{  \det{M^{( {\rm{W}} )}}
      }{
         \det{M^{(    0     )}}
      }
  =
  1 
  + 
  \sum_{i=1}^{2N^2}{
         \frac{ M^{({\rm{int}})}_{ii}
              }{
                \lambda_i^{(0)}
              }
        }
  -\frac{1}{2}
   \sum_{i,j=1}^{2N^2}{
              \frac{ M^{({\rm{int}})}_{ij} 
                     M^{({\rm{int}})}_{ji}
                   }{
                     \lambda_i^{(0)}
                     \lambda_j^{(0)}
                    }
             }
 +\frac{1}{2}
  \sum_{i,j=1}^{2N^2}{
              \frac{ M^{({\rm{int}})}_{ii} 
                     M^{({\rm{int}})}_{jj}
                   }{
                     \lambda_i^{(0)}
                     \lambda_j^{(0)}
                    }
             }
+ \dots
\quad .
\label{dia}
\end{equation}
Here the indices $i$ and $j$ stand for the combination of Dirac
index and momenta $ ( \alpha, p ) $ which 
label fermionic matrix elements, so that $M_{ij}^{({\rm{int}})}
\equiv M_{(\alpha p)(\beta q)}^{({\rm{int}})}$
represents $\langle \lambda_\alpha^{(0)} (p) | M^{({\rm{int}})}(p,q)
| \lambda_\beta^{(0)} (q) \rangle $.
The traces in (\ref{dia}) are, in fact,
the diagrams which contribute to
the vacuum polarization tensor.

Setting
\begin{eqnarray}
 t_i & = &  \langle M_{ii}^{({\rm{int}})} \rangle 
\quad, 
\label{s1}
\\
 s_{ij} & = & s_{ji} =
\langle M_{ii}^{({\rm{int}})} M_{jj}^{({\rm{int}})}\rangle 
\quad ,
\label{t1}
\\
 t_{ij} & = & t_{ji} 
= \langle M_{ij}^{({\rm{int}})} M_{ji}^{({\rm{int}})}\rangle 
-
 s_{ij}
\label{t2}
\quad ,
\end{eqnarray} 
the ratio of the interactive and free 
partition functions may be written
\begin{equation}
\frac{ 
  \left\langle     \det{M^{({\rm{W}})}   } \right\rangle 
      }{
       \det{M^{(0)}}
      }
 = 
 1
 +
 \sum_{i=1}^{2N^2}{
                   \frac{t_i}{\lambda_i^{(0)}}
                  } 
-\frac{1}{2}  
 \sum_{i,j=1}^{2N^2}{
                      \frac{
                            t_{ij}
                          }{
                            \lambda_i^{(0)}
                            \lambda_j^{(0)}
                           } 
                     }
 + \dots
\quad  .
\label{additive}
\end{equation}
This Taylor expansion is analytic in $1/2\kappa$ 
with poles at $\lambda_i^{(0)}=0$ or $1/2\kappa = \eta_i^{(0)}$. 

The Wilson fermion matrix  $M^{({\rm{W}})}$ is a $2N^2$ dimensional 
square matrix given by (\ref{3.2})-(\ref{3.12}).
Its determinant, and the bosonic expectation value thereof,
are therefore  polynomials of degree $2N^2$
with corresponding number of zeroes. As such, 
the latter  may be written
(up to an irrelevant constant)
\begin{equation}
     \left\langle 
      \det{M^{({\rm{W}})}}
 \right\rangle 
= 
 \prod_{i=1}^{2N^2}
\left( 1/2\kappa - \eta_i \right)
\quad ,
\label{multiplicative1}
\end{equation}
where $\eta_i$ represents $\eta_\alpha(p)$ and are the Lee-Yang zeroes
of the interactive model. These are the quantities to be determined
at weak coupling.

Writing 
\begin{equation}
\Delta_i = {\eta}_i - {\eta_i}^{(0)}
\quad ,
\label{sft}
\end{equation}
gives, now, a new 
type of weak coupling expansion for the ratio of partitions
functions, which 
is `multiplicative' rather than additive in form,
\begin{equation}
 \frac{ 
     \left\langle 
               \det{M^{({\rm{W}})}}
     \right\rangle 
      }{
       \det{M^{(0)}}
      }
= 
 \prod_{i=1}^{2N^2}
\left( \frac{1/2\kappa - \eta_i}{\lambda_i^{(0)}} \right)
 = 
\prod_{i=1}^{2N^2}\left(
  1 - \frac{\Delta_i}{1/2\kappa - \eta_i^{(0)}}
                    \right)
 \quad .
\label{multiplicative}
\end{equation}
Note that the expression (\ref{multiplicative}),
like its additive counterpart, (\ref{additive}),
analytic in $1/2\kappa$ with
poles at $\eta_i^{(0)}$.

Expanding (\ref{multiplicative}) gives
\begin{equation}
 \frac{ 
     \left\langle 
      \det{M^{({\rm{W}})}}
 \right\rangle 
      }{
       \det{M^{(0)}}
      }
= 1 -
 \sum_{i=1}^{2N^2}{
 \frac{\Delta_i}{\lambda_i^{(0)}}
}
+
\frac{1}{2} \sum_{i=1}^{2N^2}{\sum_{j\ne i}^{2N^2}{
 \frac{\Delta_i \Delta_j}{\lambda_i^{(0)}\lambda_j^{(0)}}
}}
+ \dots
\quad.
\label{expmult}
\end{equation}

Let $\{n\}$ denote the $n^{\rm{th}}$ 
degeneracy class in the free fermion case, 
so that the $D_n$ eigenvalues
$\lambda^{(0)}_{n_1} = \dots
 =\lambda^{(0)}_{n_{D_n}}$ are identical to $\lambda^{(0)}_{n}$, say,
with $D_n=2$ or $4$.
Take the hopping parameter to be complex and arbitrarily close
to a free fermion zero,
\begin{equation}
1/2\kappa = \eta_n^{(0)} + \epsilon
\quad .
\end{equation}
The additive and multiplicative expressions (\ref{additive}) and
(\ref{expmult}) for the ratio of partition functions may
now be expanded in $\epsilon^{-1}$.
Indeed, (\ref{additive}) gives
\begin{equation}
\frac{ 
  \left\langle     \det{M^{({\rm{W}})}   } \right\rangle 
      }{
       \det{M^{(0)}}
      }
 = 
 - \epsilon^{-2}
 \frac{1}{2}
 \sum_{n_i,n_j\in \{n\}}{t_{n_i n_j}}
 + \epsilon^{-1}
 \left\{
 \sum_{n_i\in \{n\}}{t_{n_i}}
 -
  \sum_{n_i\in \{n\},j\not\in \{n\}}{
\frac{t_{n_i j}}{\eta_n^{(0)}-\eta_j^{(0)}+\epsilon}
}
\right\}
+ {\cal{O}}\left(\epsilon^0\right)
\quad  ,
\label{additiveexp}
\end{equation}
while (\ref{expmult}) yields
\begin{eqnarray}
\frac{ 
  \left\langle     \det{M^{({\rm{W}})}   } \right\rangle 
      }{
       \det{M^{(0)}}
      }
&  = & 
  \epsilon^{-2}
 \sum_{n_i,n_j\in \{n\},n_i \ne n_j}{
 \frac{\Delta_{n_i}\Delta_{n_j}}{2}
}
\nonumber \\
& & 
 + \epsilon^{-1}
 \left\{
- \sum_{n_i\in \{n\}}{\Delta_{n_i}}
 +
  \sum_{n_i\in \{n\},j\not\in \{n\}}{
\frac{\Delta_{n_i} \Delta_j}{\eta_n^{(0)}-\eta_j^{(0)}+\epsilon}
}
\right\}
+ {\cal{O}}\left(\epsilon^0\right)
\quad  .
\label{multexp}
\end{eqnarray}
Equating these two expansions to ${\cal{O}}(\epsilon^{-2})$ gives
\begin{equation}
 \sum_{n_i,n_j\in \{n\},n_i \ne n_j}{\Delta_{n_i} \Delta_{n_j}}
=
- \sum_{n_i,n_j\in \{n\}}{t_{n_i n_j}}
\quad ,
\label{entire}
\end{equation}
while to ${\cal{O}}(\epsilon^{-1})$ it gives
\begin{equation}
 \sum_{n_i\in \{n\}}{
 \Delta_{n_i}
 \left\{
         1 -  \sum_{j\not\in \{n\}}{
                         \frac{\Delta_j}{\eta_n^{(0)}-\eta_j^{(0)}}
                                   }
 \right\}
 }
=
  \sum_{n_i\in \{n\},j\not\in \{n\}}{
\frac{t_{n_i j}}{\eta_n^{(0)}-\eta_j^{(0)}}
}
-
\sum_{n_i\in \{n\}}{t_{n_i}}
\quad ,
\end{equation}
having taken the $\epsilon \rightarrow 0$ limit.

Let
\begin{eqnarray}
 t_i & = & t_i^{(1)} + t_i^{(2)} + {\cal{O}}({{g}}^3) \quad,
\\
 t_{ij} & = &  t_{ij}^{(2)} + {\cal{O}}({{g}}^3) \quad,
\\
\Delta_i & = & {\eta}_i^{(1)} + {\eta}_i^{(2)}
+ {\cal{O}}({{g}}^3) 
\quad ,
\end{eqnarray}
 where 
$t_i^{(1)}$  and ${\eta}_i^{(1)}$
are the 
order ${{g}}$ contributions to the expectation values of the matrix 
elements and zero shifts and where
$t_i^{(2)}$, $t_{ij}^{(2)}$ and ${\eta}_i^{(2)}$
are their 
order ${{g^2}}$ equivalents.
The ${\cal{O}}(\epsilon^{-1})$ equation to order ${g}$
is
\begin{equation}
 \sum_{n_i \in \{n\}}{ {\eta}_{n_i}^{(1)} }  = 
 -
 \sum_{n_i \in \{n\}}{ t_{n_i}^{(1)} } 
\quad ,
\label{M27.1}
\end{equation}
and its order ${g}^2$ counterpart  is
\begin{equation}
  \sum_{n_i \in \{n\}}{ {\eta}_{n_i}^{(2)} }  =  
-  \sum_{n_i \in \{n\}}{ t_{n_i}^{(2)} } 
+  \sum_{n_i \in \{n\},j \not\in \{n\} }{
 \frac{t_{n_ij}^{(2)} +t_{n_i}^{(1)}t_j^{(1)}
  }{\eta_n^{(0)}-\eta_j^{(0)}}
}
\label{M27.2}
\quad .
\end{equation}
Also, the ${\cal{O}}(\epsilon^{-2})$ equation, (\ref{entire}), is, now,
\begin{equation}
 \sum_{
        n_i \in \{n\}
      }{
             \left(
               {\eta_{n_i}}^{(1)}
         \right)^2
      } 
  = 
 \sum_{
       n_i,n_j \in \{n\}
      }{
                 t_{n_i n_j}^{(2)}
       }
+
\left( \sum_{
       n_i \in \{n\}
      }{
                 t_{n_i}^{(1)}
       }
\right)^2
\label{M27.3}
\quad .
\end{equation}
With relations (\ref{M27.1})-(\ref{M27.3}), 
the multiplicative expression (\ref{expmult}) recovers (\ref{additive})
to ${\cal{O}}({g}^{2})$. Thus, equating
the ${\cal{O}}\left(\epsilon^0\right)$
contributions to  (\ref{additiveexp}) and (\ref{multexp}) yields no
extra information.

The partition function zeroes are `protocritical points'
in the sense that they have the potential to become
true critical points \cite{proto}. In the limit of infinite volume, 
the lowest zeroes
 impact on to the real hopping parameter axis precipitating
the phase transition.
The real parts of the lowest zeroes are therefore pseudocritical 
points in the statistical mechanics sense.

In the free case, the lowest 
zeroes, and those  responsible for criticality, are two
fold degenerate. One expects critical behaviour in 
the weakly coupled case to be governed by their
equivalents there.
The two equations, (\ref{M27.1}) and (\ref{M27.3}), allow
full determination of the first order shifts to two-fold 
degenerate zeroes. Indeed, 
\begin{equation}
 \eta^{(1)}_{n_i} =
 \frac{1}{2}
 \left\{
         -t_{n_1}^{(1)}
         -t_{n_2}^{(1)}
         \pm
 \sqrt{
       (
          t_{n_1}^{(1)}
          +
          t_{n_2}^{(1)}
       )^2
       + 
         4t^{(2)}_{n_1 n_2}
      }
 \right\}
\quad ,
\label{1st}
\end{equation}
where $n_i \in \{ n \}$ for $i =1$ or $2$.
The second order equation, (\ref{M27.2}),  
in the two-fold degenerate case is
\begin{equation}
\eta^{(2)}_{n_1}
+  \eta^{(2)}_{n_2}
 = 
 -t_{n_1}^{(2)}
 -t_{n_2}^{(2)}
+
\sum_{j\not\in \{n\}}{
                \frac{
 (t_{n_1}^{(1)}
 +t_{n_2}^{(1)})t_j^{(1)}
                           + t^{(2)}_{jn_1}
                           + t^{(2)}_{jn_2}
                     }
                     {
     \eta_n^{(0)} - \eta_j^{(0)}
                     }
                      }
\quad .
\label{av}
\end{equation}
To find the individual shifts, let
\begin{eqnarray}
 \eta^{(2)}_{n_1} & = & \eta^{(2)}_{n} + \delta^{(2)}
\quad ,
\\
 \eta^{(2)}_{n_2} & = & \eta^{(2)}_{n} - \delta^{(2)}
\quad .
\end{eqnarray}
Their average, $\eta^{(2)}_{n}$, is determined directly from
(\ref{av}).
Removing the expectation values over the bosonic fields
converts the zeroes to the shifts in the eigenvalues of the
fermion matrix in the presence of a small perturbation,
$M^{\rm{(int)}}$.
The problem of determining such shifts is simply 
(two-fold degenerate) time independent
perturbation theory.
Indeed, one finds, for example,
\begin{eqnarray}
 \lambda_{n_1}
& = &
 \lambda_{n}^{(0)}
 -
 \frac{1}{2}
 \left\{
         M^{\rm{(int)}}_{n_1 n_1}
       + M^{\rm{(int)}}_{n_2 n_2}
         \pm
 \sqrt{
(M^{\rm{(int)}}_{n_1 n_1}- M^{\rm{(int)}}_{n_1 n_1})^2 
+ 4 M^{\rm{(int)}}_{n_1 n_2} M^{\rm{(int)}}_{n_2 n_1}
}
 \right\}
\nonumber \\
& & 
+
\frac{1}{2}
\sum_{j \not\in\{n\}}{
\frac{
 M^{\rm{(int)}}_{j n_1}  M^{\rm{(int)}}_{n_1 j}
+
 M^{\rm{(int)}}_{j n_2}  M^{\rm{(int)}}_{n_2 j}
}{\lambda_{n}^{(0)}-\lambda_{j}^{(0)}}
}
-
\delta_0^{(2)}
\quad ,
\label{tipt}
\end{eqnarray}
in which $\left\langle \delta_0^{(2)} \right\rangle = \delta^{(2)}$.
This recovers  time independent
perturbation theory if
\begin{equation}
 \delta_0^{(2)}
 =
 -\frac{1}{2}
 \sum_{j \not\in\{n\}}{
\frac{
 M^{\rm{(int)}}_{j n_1}  M^{\rm{(int)}}_{n_1 j}
-
 M^{\rm{(int)}}_{j n_2}  M^{\rm{(int)}}_{n_2 j}
}{\lambda_{n}^{(0)}-\lambda_{j}^{(0)}}
}
\quad ,
\end{equation}
whence
\begin{equation}
 \delta^{(2)}
 =
  \frac{1}{2}
 \sum_{j \not\in\{n\}}{
\frac{
 t^{(2)}_{j n_1}+s^{(2)}_{j n_1}
 -
 t^{(2)}_{j n_2}-s^{(2)}_{j n_2}
}{\eta_{n}^{(0)}-\eta_{j}^{(0)}}
}
\quad .
\end{equation}
Finally, the full expression for the second order shift in 
an erstwhile two-fold degenerate zero is
\begin{equation}
  \eta^{(2)}_{n_i}
= 
-\frac{1}{2} \left(t_{n_1}^{(2)} +  t_{n_2}^{(2)}\right)
+
\frac{1}{2} \
\sum_{j\not\in \{n\}}{
                      \frac{
                            2( t_{jn_i}^{(2)}+s_{jn_i}^{(2)})
 +
  t_{j}^{(1)}t_{n_1}^{(1)}
  +
  t_{j}^{(1)}t_{n_2}^{(1)}
 -
 s_{jn_1}^{(2)}-s_{jn_2}^{(2)}
                           }{
               \eta_n^{(0)} - \eta_j^{(0)}
                            }
                      }
\quad .
\label{2nd}
\end{equation}

\section{The Zeroes and Phase Diagram  of the Gross-Neveu Model}
\setcounter{equation}{0}

The interactive part of the fermion matrix (\ref{3.12})
may be split into
\begin{equation}
 M^{\rm{(int)}}(q,p) 
=
 M^{\rm{(int)}}_\phi(q,p) 
+
 M^{\rm{(int)}}_\pi(q,p) 
\quad ,
\label{GN5.1}
\end{equation}
where
\begin{eqnarray}
 M^{\rm{(int)}}_\phi(q,p) & = & 
\frac{1}{N^2}
\sum_n{
       e^{i (p-q) n a}
       \phi(n)
}
=\left(\frac{1}{Na}  \right)^2
{\phi}(q-p)
\quad ,
\label{GN5.2}
\\
 M^{ \rm{ (int) } }_\pi(q,p) & = & 
\frac{1}{N^2}
\sum_n{
        e^{i (p-q) n a}
        \pi (n) i \gamma_S
}
=\left(\frac{1}{Na}  \right)^2
\pi (q-p) i \gamma_5
\quad .
\label{GN5.3}
\end{eqnarray}
One notes that the momentum dependency of the bosonic field variables
involves even integers, so  the bosons have periodic
boundary conditions.
The generic matrix elements required for the calculation of 
(\ref{s1}), (\ref{t1}) and  (\ref{t2}) are
\begin{eqnarray}
 {M^{\rm{(int)}}_\phi}_{(\alpha p)(\beta q)}
 & = & 
 \left(\frac{1}{Na}\right)^2
 {\phi}(p-q)
 \frac{1}{2}
\times
\nonumber \\
& &  \left(
        1 + (-1)^{\alpha + \beta }
 \frac{\sum_\mu{\sin{p_\mu a}\sin{q_\mu a} 
                + i (\sin{p_1 a}\sin{q_2 a}
                    -\sin{p_2 a}\sin{q_1 a}
                    )}
}{
\sqrt{\sum_\mu{\sin^2{p_\mu a} }}
\sqrt{\sum_\mu{\sin^2{p_\mu a} }}
}
 \right)
,
\\
 {M^{\rm{(int)}}_\pi}_{(\alpha p)(\beta q)}
 & = & 
 \left(\frac{1}{Na}\right)^2
 {\pi}(p-q)
 \frac{i}{2}
\times
\nonumber \\
& &  \left(
        1 - (-1)^{\alpha + \beta }
 \frac{\sum_\mu{\sin{p_\mu a}\sin{q_\mu a} 
                + i (\sin{p_1 a}\sin{q_2 a}
                    -\sin{p_2 a}\sin{q_1 a}
                    )}
}{
\sqrt{\sum_\mu{\sin^2{p_\mu a} }}
\sqrt{\sum_\mu{\sin^2{p_\mu a} }}
}
 \right).
\end{eqnarray}

In the (generalized) Gross-Neveu case, the pure bosonic action is given
by (\ref{pure}). The pure bosonic expectation values 
 in momentum space are thus
\begin{eqnarray}
\langle \phi (k) \rangle &  = & 
\langle \pi  (k) \rangle    = 0
\quad ,
\\
 \langle \phi (k) \phi(-k)\rangle & = &
N^d a^{2d} 2 g_\phi^2
\quad ,
\\
\langle \pi(k) \pi (-k) \rangle 
\label{phimom}
&=&
N^d a^{2d} 2 
g_\pi^2 
\quad . 
\label{pimom}
\end{eqnarray}

The  bosonic  expectation values of the matrix elements 
required in the calculation of the shifts (\ref{1st}) 
and (\ref{2nd}) are then
\begin{eqnarray}
 t_i & \equiv & t_{\alpha,p} = 0
\quad ,
\label{a} 
\\
 s_{ij} & \equiv  & s_{(\alpha, p) (\beta,q) }
 =
 \frac{2{g_\phi}^2}{N^2}
\quad ,
\label{b} 
\\
 t_{ij} & \equiv  & t_{(\alpha, p) (\beta,q) } 
 = \frac{{g_\phi}^2+{g_\pi}^2}{N^2}
\left\{
  (-1)^{\alpha+\beta}
 \frac{\sum_\rho 
                \sin{p_\rho}
                \sin{q_\rho}
     }{
       \sqrt{
             \sum_\mu
                      \sin^2{p_\mu}
             \sum_\nu
                      \sin^2{q_\nu}
            }
       }
 -1
\right\}
\quad .
\label{c} 
\end{eqnarray}
From  these equations, together with (\ref{1st}) 
and (\ref{2nd}),
the ${\cal{O}}({g})$ and ${\cal{O}}({g}^2)$   
shifts
 for the erstwhile two-fold degenerate zeroes,
$\eta_\alpha(\pm|p_1|,p_2)$ (for $\hat{p}_2= 0$ or
 $-N/2$), are, respectively,
\begin{eqnarray}
 \eta^{(1)}_\alpha(\pm|p_1|,p_2)
 & = &
 \pm
 i \frac{\sqrt{2}\sqrt{g_\phi^2+g_\pi^2}}{N}
\quad ,
\label{1c}
\\
 \eta^{(2)}_\alpha(\pm|p_1|,p_2)
 & = &
- \frac{g_\phi^2+g_\pi^2}{N^2}
\sum_{(\beta,q)\not\in\{(\alpha,p)\}}{\frac{1}{\eta^{(0)}_\alpha(p)
-\eta^{(0)}_\beta(q)
}}
\quad .
\label{2c}
\end{eqnarray}
So the two four-fermi interactions in fact contribute the same amounts
to the shifts in the zeroes.

 In the thermodynamic limit these  lowest zeroes
become the true
critical points of the theory and their determination amounts to 
determination of the 
phase diagram. I.e., 
the phase diagram is given to order ${{g}}^2$ by the 
limit
\begin{equation}
\frac{1}{2\kappa(g)} = \lim_{N\rightarrow \infty}{\left\{
\eta_\alpha^{(0)}(p)
+
\eta_\alpha^{(1)}(p)
+
\eta_\alpha^{(2)}(p)
\right\}}
\quad ,
\end{equation}
where $p$ is the momentum corresponding to the lowest zeroes.
The first order shift in  (\ref{1c})  gives the relative separation
of the erstwhile two-fold degenerate zeroes and vanishes
in the infinite volume limit.
The average shift is represented by (\ref{2c}) and is second order.
The shift in the corresponding critical point is
\begin{equation}
 \lim_{N \rightarrow \infty}{
 \eta_\alpha^{(2)}(\pm|p_1|,p_2)
}
=
-
(g_\phi^2+g_\pi^2)
 \lim_{N \rightarrow \infty}{c(N)}
\quad ,
\label{soln}
\end{equation}
where
\begin{equation}
 c(N)=
\frac{1}{N^2}
\sum_{(\beta,q)\not\in\{(\alpha,p)\}}{\frac{1}{\eta^{(0)}_\alpha(p)
-\eta^{(0)}_\beta(q)
}}\quad .
\label{factor}
\end{equation}
One finds, numerically, that the imaginary contribution to
this factor
vanishes in the thermodynamic limit, meaning that these
zeroes indeed impact on to the real hopping parameter axis.
The real part of (\ref{factor}) becomes an $N$-independent constant
whose actual value depends on the free zero from which it evolved.
Indeed, (\ref{factor})
approaches approximately $0.77$ and $-0.77$ for
$(\hat{p}_1,\hat{p}_2) = (\pm1/2,0)$
and 
$(\pm(N-1)/2,-N/2)$ respectively.
These correspond to the rightmost and leftmost critical lines
(see Fig.~1).
Also, (\ref{factor}) is
approximately $0.2$ and $-0.2$ for
$(\hat{p}_1,\hat{p}_2) = (\pm(N-1)/2,0)$
and 
$(\pm 1/2,-N/2)$ respectively. These give the  two lines that
generate
the inner cusp.
The situation is summarized in Table~\ref{taba} where
the critical
hopping parameters in the free and interacting  cases 
and the  momentum
indices of the corresponding zeroes are listed.
\noindent
\begin{table}[ht]
\caption{The critical points and the momentum indices of their 
corresponding
partition function zeroes in the free and weakly interacting cases.
}
\label{taba}
\begin{center}
\begin{small}
\vspace{0.5cm}  
\noindent\begin{tabular}{|c|c|c|c|c|} 
    \hline 
    \hline
      & & &  & \\
     $\left( \hat{p_1}, \hat{p}_2 \right)$ &
     $\left( \pm \frac{1}{2}(N-1), -\frac{N}{2} \right)$ &
     $\left( \pm \frac{1}{2}, -\frac{N}{2}  \right)$ &
     $\left( \pm \frac{1}{2}(N-1), 0 \right)$ &
     $\left( \pm \frac{1}{2}, 0 \right)$ \\
      & & &  & \\
    \hline
      & & &  & \\
     $1/2\kappa_c(0)$ &
     $-2$ &
     $0$ &
     $0$ &
     $2$ \\
      & & &  & \\
    \hline
      & & &  & \\
     $1/2\kappa_c(g)$ &
     $-2+0.77 (g_\phi^2+g_\pi^2)$ &
     $0.2(g_\phi^2+g_\pi^2)$ &
     $-0.2(g_\phi^2+g_\pi^2)$ &
     $2-0.77 (g_\phi^2+g_\pi^2)$ \\
      & & &  & \\
    \hline
    \hline
  \end{tabular}
\end{small}
\end{center}
\end{table}

\begin{figure}[htb]
\vspace{6.5cm}               
\includegraphics{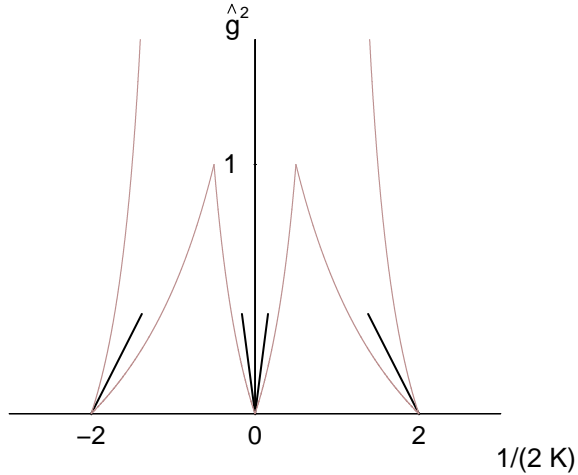}
\caption{The phase diagram for the
Gross-Neveu model 
in the weakly coupled region $g_\phi = g_\pi = g$ 
(to ${\cal{O}}({g}^2)$)
(dark lines) and  a schematic representation of
the expected Aoki phase diagram (light curves).}
\label{fig}
\end{figure}  
The actual phase diagram for weakly coupling
is pictured in Fig.~1 for $g_\phi = g_\pi = g$ (dark lines). 
The lighter curves are 
a schematic representation of
the expected full phase diagram.
One sees the degeneracy of the central 
free fermion critical point 
is lifted and two critical lines emerge in the
presence of weak bosonic coupling.
These are the
lines corresponding to the central  cusp in Aoki's phase diagram.
From (\ref{soln}), the central cusp can only be made vanish at this
order, in the unphysical situation of imaginary couplings.
The Aoki phase does not yet emerge to ${\cal{O}}({{g}}^2)$
from the left- and rightmost free critical points.

In the free case, the zeroes and hence the critical points
are invariant under momentum inversion $p_\mu \rightarrow -p_\mu$,
corresponding to a rotation in space-time. While this
degeneracy is lifted at finite size in the interacting
case, it is recovered in the limit of infinite volume.
In that limit, the free zeroes and critical points are 
also invariant under 
the parity transformation $p_1 \leftrightarrow p_2$. This is
no longer the case in the presence of interactions. 
Indeed, the inner pair of critical lines are interchanged
under parity.
The overall phase structure, however,  remains the same.

The situation is similar to the two dimensional Potts model. 
There, the partition function is invariant under a duality
transformation which exchanges the high and low temperature phases.
The critical point is that which is invariant under that transformation.
Here, the zeroes, and hence the partition function, are invariant under
parity. The phase structure is also unchanged by parity.
However, parity even and parity odd regions of the phase diagram
are interchanged.

\section{Conclusions}
\setcounter{equation}{0}

A new type of weak coupling expansion appropriate for
Wilson fermionic lattice field theories  has been developed.
This expansion is multiplicative, but  recovers
the standard additive expansion.
Its multiplicative form allows the Lee-Yang zeroes
of the weakly coupled theory to be extracted in a natural way.
These zeroes, are protocritical points, which, if they
impact on to the real hopping parameter axis, precipitate a
phase transition there.

The expansion is applied to the single flavour
lattice Gross-Neveu model
to track the movement of zeroes and thereby the critical points 
in the presence of bosonic field variables.
This model shares features with QCD, one of which is expected to
be the existence of an Aoki phase.

Using the new weak coupling expansion,
a  phase diagram is obtained in the weakly coupled 
region which is consistent with
that of Aoki. The widths of 
the Aoki cusps are analytically determined to second order
in the couplings.
The central cusp cannot be tuned away for real physical couplings. 
The lateral cusps do not yet emerge at this order.
This is the answer to the question posed by Creutz in \cite{Creutz}
for the single flavour Gross-Neveu model.

Finally, while the full phase structure is unaltered by a parity 
transformation, such an operation has the effect of exchanging
the critical lines forming the inner Aoki cusp.


\end{document}